\documentclass[sigconf]{acmart}
\usepackage{graphicx} 
\usepackage{float}
\usepackage{booktabs} 
\usepackage{enumitem,kantlipsum}
\usepackage{threeparttable}
\usepackage{algorithm}
\usepackage{algpseudocode}
\usepackage{multirow}
\usepackage{array}
\usepackage{booktabs}
\usepackage{subfig}
\usepackage{pgfplots}
\usepackage{tikz}
\usepgfplotslibrary{groupplots}

\usepackage[textsize=tiny]{todonotes}

\copyrightyear{2022}
\acmYear{2022}
\setcopyright{acmlicensed}\acmConference[SIGIR '22]{Proceedings of the 45th
International ACM SIGIR Conference on Research and Development in
Information Retrieval}{July 11--15, 2022}{Madrid, Spain}
\acmBooktitle{Proceedings of the 45th International ACM SIGIR Conference on
Research and Development in Information Retrieval (SIGIR '22), July 11--15,
2022, Madrid, Spain}
\acmPrice{15.00}
\acmDOI{10.1145/3477495.3531714}
\acmISBN{978-1-4503-8732-3/22/07}

\begin{document}
\fancyhead{}

\title{Rethinking Reinforcement Learning for Recommendation:\\A Prompt Perspective}

\author{Xin Xin}
\affiliation{%
  \institution{Shandong University}
  \country{China}
}
\email{xinxin@sdu.edu.cn}

\author{Tiago Pimentel}
\affiliation{%
  \institution{University of Cambridge}
  \country{United Kingdom}
}
\email{tp472@cam.ac.uk}

\author{Alexandros Karatzoglou}
\affiliation{%
  \institution{Google Research}
  \country{United Kingdom}
}
\email{alexkz@google.com}

\author{Pengjie	Ren}
\affiliation{%
  \institution{Shandong University}
  \country{China}
}
\email{jay.ren@outlook.com}

\author{Konstantina	Christakopoulou}
\affiliation{%
  \institution{Google}
  \country{United States}
}
\email{konchris@google.com}

\author{Zhaochun Ren}
\affiliation{%
  \institution{Shandong University}
  \country{China}
}
\email{zhaochun.ren@sdu.edu.cn}

\begin{abstract}

Modern recommender systems aim to improve user experience. As reinforcement learning (RL) naturally fits this objective---maximizing an user's reward per session---it has become an emerging topic in recommender systems. Developing RL-based recommendation methods, however, is not trivial due to the \emph{offline training challenge}. Specifically, the keystone of traditional RL is to train an agent with large amounts of online exploration making lots of `errors' in the process. In the recommendation setting, though, we cannot afford the price of making `errors' online. As a result, the agent needs to be trained through offline historical implicit feedback, collected under different recommendation policies; traditional RL algorithms may lead to sub-optimal policies under these offline training settings. 

Here we propose a new learning paradigm---namely Prompt-Based Reinforcement Learning (PRL)---for the offline training of RL-based recommendation agents. While traditional RL algorithms attempt to map state-action input pairs to their expected rewards (e.g., Q-values), PRL directly infers actions (i.e., recommended items) from state-reward inputs. In short, the agents are trained to predict a recommended item given the prior interactions and an observed reward value---with simple supervised learning. At deployment time, this historical (training) data acts as a knowledge base, while the state-reward pairs are used as a prompt. The agents are thus used to answer the question: \emph{ Which item should be recommended given the prior interactions \& the prompted reward value}? We implement PRL with four notable recommendation models and conduct experiments on two real-world e-commerce datasets. Experimental results demonstrate the superior performance of our proposed methods.

\end{abstract}
%
%
\ccsdesc[500]{Information systems~Recommender systems}
\ccsdesc[500]{Information systems~Retrieval models and ranking}
\ccsdesc[500]{Information systems~Novelty in information retrieval}

\keywords{Next Item Recommendation; Reinforcement Learning; Recommender Systems; Session-based Recommendation}

\maketitle

\section{Introduction}
\begin{figure*}
    \captionsetup[subfloat]
    {}
    \centering
    \subfloat[Policy evaluation.]{%
    \label{fig:pe}
    \includegraphics[width=0.31\textwidth]{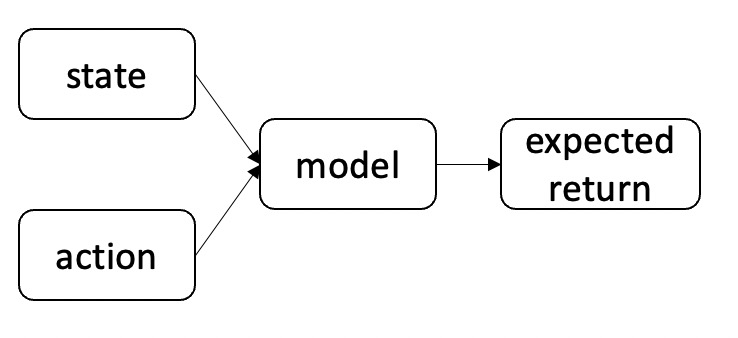}}
    \hspace{0.2cm}
    \subfloat[Policy improvement.]{%
    \label{fig:pi}
    \includegraphics[width=0.31\textwidth]{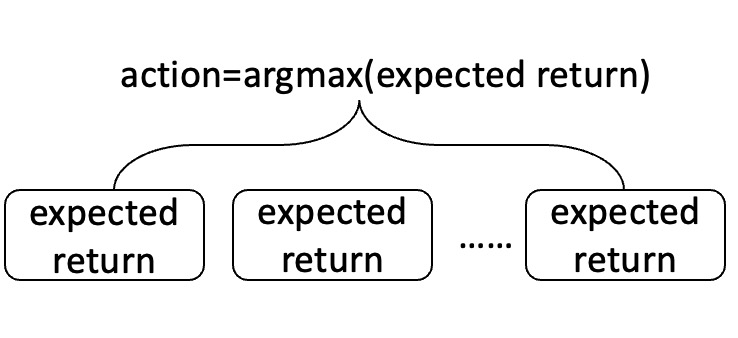}}
    \hspace{0.2cm}
    \subfloat[PRL.]{
    \label{fig:prl}
    \includegraphics[width=0.31\textwidth]{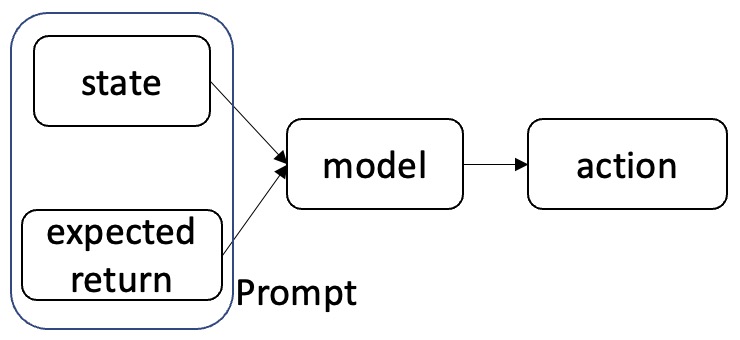}}
    \caption{Traditional RL algorithms involve policy evaluation (a) to predict the expected return (i.e., cumulative reward) and then use policy improvement (b) to select actions with the highest return prediction. While PRL (c) aims to directly infer actions given the prompt of current state and expected return.}
\end{figure*}
Next item  recommender  systems  are  one  of  the  core components  of  several  modern  online  web  services,  including  music or video streaming  services~\cite{nextitnet},  and  e-commerce  sites~\cite{reinforce-e-commerce}.   They are  holistically  ingrained  into  these  applications,  helping users navigate and find new content. 
As a general rule, these systems are modelled as sequence prediction tasks—they thus answer the question:{\it What is the next item the user would be interested to interact with given the past interactions}—and  are  typically implemented on top of recurrent neural networks or other generative  sequential  models.
 Conventional next item recommendation models
are usually trained through an auto-regressive fashion, in which the model is trained to recover the historical interaction sequence \cite{gru4rec,nextitnet,SASRec}. Similar learning objectives are also used in language modeling in the field of natural language processing (NLP).
Simply predicting the next item a user will interact with, however, may be a poor objective; one might prefer to instead maximise long-term engagement, for instance, or the diversity of the consumed items.

Reinforcement learning (RL) has been successfully employed in planning and controlling \cite{alphogo,humanlevelcontrol}. An RL agent is trained to take actions which, given the observed state of the environment, maximize a pre-defined reward. Existing value-based RL algorithms usually involve policy evaluation and policy improvement, as shown in Figures \ref{fig:pe} and \ref{fig:pi}, respectively. Policy evaluation aims to learn a model which maps the state-action input pairs to the expected cumulative rewards (i.e., Q-values). Policy improvement selects the action with the maximum Q-value prediction. The long-term nature of planning in RL fits naturally with desirable properties in recommender systems (RS). The flexible reward setting in RL enables for flexible customization of  recommendation objectives. As a result, the use of RL in recommendation has become an emerging topic \cite{chen2021survey,zhao2019survey,afsar2021reinforcementsurvey}.

However, developing RL-based recommendation methods is non-trivial. The learning paradigm of RL trains the agent by interacting with the environment and then observing the reward. The agents are reinforced towards taking actions with higher cumulative returns. This process needs a large amount of interactions taken by the agent itself. Although some existing RL methods are claimed to be ``off-policy'', they still need the agent to step over plenty of online interactions to refresh the replay buffer. Such a learning paradigm is feasible in  fields like gaming \cite{alphogo}, since conducting error-prone explorations does not come at a cost. In the field of RS, however, we cannot afford the price of making errors, since bad recommendation results will definitely affect user experience. As a result, we want to train the RL recommendation agent through fixed historical data without the agent being able to probe the environment. However, this historical data is not generated by the target agent itself, but from different or even unknown behavior policies. The expected quality of an estimated policy can be easily affected by this discrepancy in distributions. This problem is know as the \emph{offline training challenge}.

Previous work attempted to address the offline training challenge through inverse propensity scores \cite{googlewsdmoffpolicycorrection}, model-based user simulation \cite{chen2019generativeusermodel}, or combining RL with supervised learning \cite{xin2020self}. However, such methods still suffer from factors of unbounded high variances \cite{munos2016safeandefficient}, biased user simulation \cite{huang2020keepingbiasout} and Q-value estimation \cite{bcq}.

We propose \emph{Prompt-Based Reinforcement Learning} (PRL), a new paradigm for effective offline training of RL-based recommendation agents.
The concept of prompting is rooted in NLP \cite{liu2021promptsurvey}, whereby large language models have been shown to learn new tasks with a single demonstration of an example. 
While PRL is not identical to this few-shot NLP concept, it is a similar concept in that we achieve the control of recommendation results by feeding the model with different prompt templates. 
PRL uses the offline historical data as a knowledge base while the state-reward pairs act as the prompt. The agents are trained to answer the question of \emph{which item should be recommended if the prompted reward value is expected to be achieved under the given state}.
In the training stage a generative sequential model is used to encode the users previous interactions into a hidden state, which can be regarded as the state of the environment in the RL setting. The current recommended item can be seen as the action.
From the offline data we can compute the exact cumulative reward at each step of an interaction session. 
As a result, the historical data can be organized in the template of \{state, cumulative reward\}-->\{observed action\}.
We then use a simple yet effective supervised self-attentive block \cite{Transformer} to learn and store such signals. 
During the inference stage, given the current state we feed the model an expected cumulative reward we want to obtain (e.g., twice the average cumulative reward of the current step), the model can directly infer actions through querying the historical knowledge base, as shown in Figure \ref{fig:prl}. PRL enables the recommendation agent to adjust its actions conditioning on the prompt reward. For example, agent exploration can be effectively achieved by introducing random noise on the prompt reward during the inference stage. 
We verify the effectiveness of our approach by implementing PRL with four renowned recommendation models.

To summarize, this work makes the following contributions:
\begin{itemize}
    \item We propose prompt-based reinforcement learning for the offline training of RL-based next item recommendation. We propose to use the state-reward pairs as the prompt to infer actions through querying the knowledge base of historical implicit feedback data.
    \item We propose to use a supervised self-attentive block to learn and store the signals between the input of state-reward pairs and the output of actions.
    \item We implement PRL with four renowned next item recommendation models as the state encoders, and conduct experiments on two real world e-commerce datasets. Experimental results demonstrate a generalized improvement of recommendation performance.
\end{itemize}
\section{Challenge Investigation}
We first formulate the task of next item recommendation. Then we introduce reinforcement learning and analyse the offline training challenge. After that, the concept of prompting is described.

\subsection{Next Item Recommendation} \label{Next-item-recommendation}

Let $\mathcal{I}$ denote the entire set of items in a specific system, then a user-item interaction sequence can be represented as $x_{1:t}=\left\{x_1,x_2,...,x_{t-1},x_t\right\}$, where $x_i \in \mathcal{I} (0< i \leq t)$ is the interacted item at timestamp $i$. 
Next item recommendation aims at recommending items that a user might be interested in at timestep $t+1$ given the sequence of past items $x_{1:t}$.
\begin{figure*}
    \captionsetup[subfloat]
    {}
    \centering
    \subfloat[On-policy optimization.]{%
    \label{fig:onpolicy}
    \includegraphics[width=0.31\textwidth]{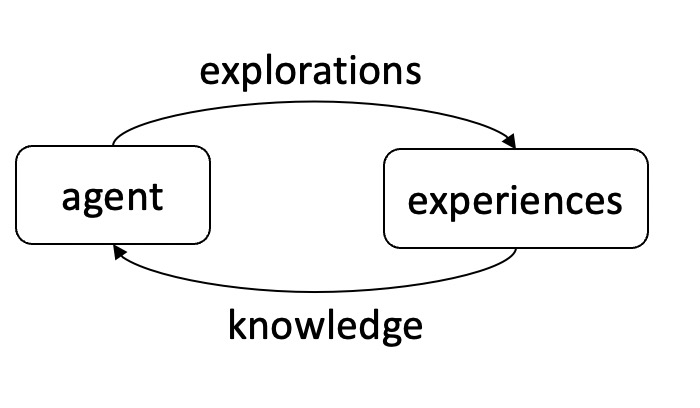}}
    \hspace{0.2cm}
    \subfloat[Off-policy optimization.]{%
    \label{fig:offpolicy}
    \includegraphics[width=0.31\textwidth]{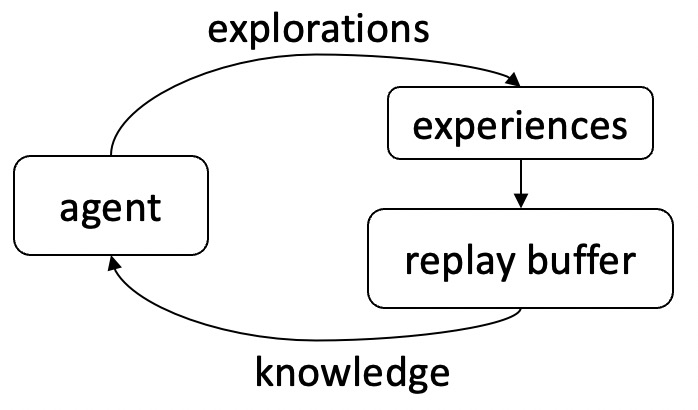}}
    \hspace{0.2cm}
    \subfloat[Offline training for RS.]{
    \label{fig:offline}
    \includegraphics[width=0.31\textwidth]{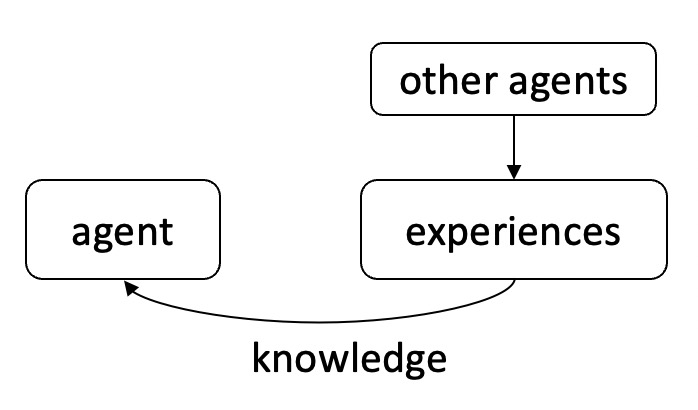}}
    \caption{On-policy optimization methods (a) need to learn from large amount of experiences taken by the agent itself. Off-policy methods (b) improve the data efficiency through introducing a replay buffer to store the past experiences. However, the stored experiences still come from the agent itself and new interactions are needed to refresh the buffer. For the offline training of RS (c), the agent is expected to be trained from experiences of other agents without new explorations and users' involvement.}
\end{figure*}

\subsection{Reinforcement Learning and the Challenge} \label{RL}

An RL agent is trained to take actions in an environment to get the maximum cumulative reward. This task is usually formulated as a Markov Decision Process (MDP) \cite{shani2005mdp,alphogo,double-q-learning}.
More precisely, for next item recommendation, users can be seen as the environment, while the MDP can be defined as tuples of $(\mathcal{S},\mathcal{A},\mathbf{P},R, \rho_0,\lambda)$ where
\begin{itemize}
    \item $\mathcal{S}$: the space of all possible user states, which can be modeled through previous item interactions.
    Concisely, we can use a sequential model $G$ to map the previous interaction sequence before timestamp $t$ into a hidden state as $\mathbf{s}_t=G(x_{1:t}) \in \mathcal{S}$ $(t>0)$. We will discuss prominent models for implementing $G(\cdot)$ in section \ref{related-work}.
    \item $\mathcal{A}$: the  discrete action space which contains candidate items. An action $a$ in the MDP represents the selection of a recommended item. In the offline training data, we can get the action at each timestamp $t$ as $a_t=x_{t+1}$. 
    \item $\mathbf{P}$: $\mathcal{S} \times \mathcal{A} \times \mathcal{S} \rightarrow \mathbb{R}$ is the state transition probability, describing how the environment state changes when an action is performed in the environment.
    \item $R$: $\mathcal{S} \times \mathcal{A} \rightarrow \mathbb{R}$ is the reward function where $r_t$ denotes the immediate reward at the $t$-th interaction step\footnote{While in general a reward is not necessarily a deterministic function of a state--action pair, we will assume so to simplify our exposition.}. This is the key component of RL, which enables the agent to be trained in a customizable reward-driven fashion, such as promoting purchases \cite{xin2020self}, increasing diversity \cite{stamenkovic2021choosing} or dwell time \cite{googlewsdmoffpolicycorrection}.
    \item $\rho_0$ describes the initial state distribution as $\mathbf{s}_0 \sim \rho_0$.
    \item $\lambda$ is the discount factor for future rewards.
\end{itemize}

The goal of RL is to find a target policy $\pi_\theta(a|\mathbf{s})$ which maps the user's state $\mathbf{s} \in \mathcal{S}$ into a probability distribution over actions $a \in \mathcal{A}$, so that if the agent samples actions according to $\pi_\theta$ the system can obtain the maximum expected cumulative reward:
\begin{equation}
	\label{cumulative-rewards}
	\max_{\pi_\theta}\mathbb{E}_{\tau\sim\pi_\theta}[R(\tau)]\text{, where }R(\tau)=\sum_{t=0}^{|\tau|}\lambda^{t}r_t(\mathbf{s}_t,a_t),
\end{equation}
where $\theta$ denotes policy parameters and  $\tau=(\mathbf{s}_0,a_0,\mathbf{s}_1,...)$ is the sampled trajectory of the target policy with $\mathbf{s}_0 \sim \rho_0$, $a_t \sim \pi_\theta(\cdot|\mathbf{s}_t)$, $s_{t+1} \sim \mathbf{P}(\cdot|\mathbf{s}_t,a_t)$.
\subsubsection{On-Policy Optimization}
On-policy optimization, e.g. policy-gradient (PG) \cite{REINFORCE}, is one of the most adopted methodologies to solve Eq.(\ref{cumulative-rewards}).  
PG aims at directly deriving the gradients of the expected cumulative rewards with respect to policy parameters $\theta$ as: \looseness=-1
\begin{equation}
	\label{policy-gradient}
	\nabla=\mathbb{E}_{\tau\sim\pi_\theta}[R(\tau)\nabla_\theta \log \pi_\theta(\tau)].
\end{equation}
Estimating this expectation requires a large amount of explorations taken by the agent itself, as shown in Figure \ref{fig:onpolicy}.
However, for the offline training of RS from historical data, all we can estimate is: 
\begin{equation}
	\label{gradient-realworld}
	\nabla'=\mathbb{E}_{\tau\sim\beta}[R(\tau)\nabla_\theta \log \pi_\theta(\tau)],
\end{equation}
where $\beta$ denotes the behavior data distribution of the historical data. 
Obviously, there is a difference between the distributions $\pi_\theta$ and $\beta$.
\cite{googlewsdmoffpolicycorrection} proposed to introduce an inverse propensity score (IPS) to correct the discrepancy at each timestamp as:
\begin{equation}
	\label{eq:correction}
	\nabla(t)=\frac{\pi_\theta(\tau(t))}{\beta(\tau(t))}\nabla'(t)\approx \frac{\pi_\theta(a_t|\mathbf{s}_t)}{\beta(a_t|\mathbf{s}_t)}\nabla'(t)
\end{equation}
However, estimating a behavior policy $\beta$ could be difficult \cite{xin2021sa2c}; and the computed IPS can have unbounded high variance \cite{afsar2021reinforcementsurvey}.

\subsubsection{Off-Policy Optimization}
Off-policy optimization methods use a replay buffer to store past experiences and improve data efficiency. Deep Q-learning \cite{alphogo} (DQN) is one of the most typical off-policy methods. DQN utilizes policy evaluation (see Figure \ref{fig:pe}) to calculate Q-values\footnote{The Q-value for a state-action pair (i.e., $Q(\mathbf{s},a)$) is defined as the expected cumulative reward gain if the action $a$ is operated under the state $\mathbf{s}$.} and then policy improvement (see Figure \ref{fig:pi}) to select actions with the highest Q-values. The model in policy evaluation is updated as follows:
\begin{equation}
\begin{aligned}
    \label{eq:tdupdate}
	\theta&\leftarrow\theta-\alpha\, \frac{\partial\, \mathbb{E}_{s,a \sim \pi_\theta'}(Q_\theta(\mathbf{s},a)-Q_T(\mathbf{s},a))^2}{\partial\theta}, \mathtt{where} \\
	&Q_T(\mathbf{s},a)=r+\lambda \max_{a'}{} \mathbb{E}_{\mathbf{s}'\sim \mathbf{P}(\cdot |\mathbf{s},a)}Q_{\theta}(\mathbf{s}',a').
\end{aligned}
\end{equation}
$\alpha$ is the learning rate, $Q_\theta(\mathbf{s},a)$ denotes the Q-value calculated from the model while $Q_T(\mathbf{s},a)$ is the target Q-value computed from time-difference (TD) learning \cite{bellman1966dynamic},
$\pi_\theta'$ denotes the policy used to build and refresh the replay buffer. 
This off-policy method requires $\pi_\theta'$ to be defined as the previous version of $\pi_\theta$ \cite{alphogo,bcq}.
In other words, the experiences stored in the replay buffer are still generated by the agent itself and new explorations are needed to refresh the replay buffer, as shown in Figure \ref{fig:offpolicy}.

However, when performing offline learning, historical training data comes from different and typically unknown agents. To avoid affecting user experience we anticipate that new explorations with user involvement are not needed until the agent is well trained, as shown in Figure \ref{fig:offline}. In such setting, what we can update is:
\begin{equation}
\begin{aligned}
    \label{eq:tdupdate_offline}
	\theta&\leftarrow\theta-\alpha\, \frac{\partial\, \mathbb{E}_{s,a \sim \beta}(Q_\theta(\mathbf{s},a)-Q'_T(\mathbf{s},a))^2}{\partial\theta}, \mathtt{where} \\
	&Q'_T(\mathbf{s},a)=r+\lambda \max_{a'}{} \mathbb{E}_{\mathbf{s}'\sim \mathbf{P}_{\beta}(\cdot |\mathbf{s},a)}Q_{\theta}(\mathbf{s}',a').
\end{aligned}
\end{equation}
where $s'\sim\mathbf{P}_{\beta}(\cdot |\mathbf{s},a)$ denotes that the next state $s'$ is sampled from the offline data, rather than operating actions online and then observe the next state. Given that the state and action distribution in $\beta$ can be different from the target policy, the parameter update in Eq.(\ref{eq:tdupdate_offline}) can easily be biased. \cite{bcq,xin2020self} have shown that off-policy methods suffer from weak performance in the offline training setting.
\begin{figure}
    \centering
    \includegraphics[width=0.45\textwidth]{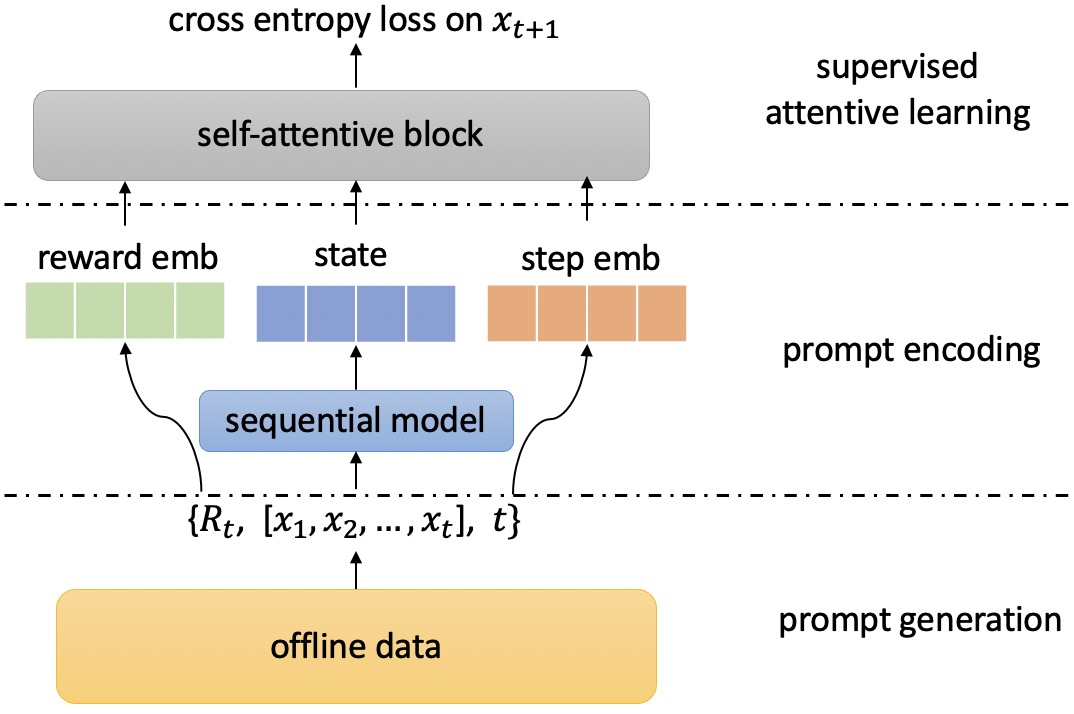}
    \caption{The training framework of PRL. ``emb'' is short for embedding. Prompt generation transforms the offline historical training data into tuples: \{cumulative reward $R_t$, previous interactions $x_{1:t}$, interaction step $t$\}. Then prompt encoding encodes the tuples into latent representations. Finally, a self-attentive block is used to learn the mapping function between the prompt and the action. The model parameters can be updated through a simple supervised cross-entropy loss function.}
    \label{fig:prltrain}
\end{figure}
\subsubsection{Model-based RL}
An alternative approach to train an RL recommendation agent is to use model-based reinforcement learning \cite{chen2019generativeusermodel,GAIL}. Model-based RL is based on a model of the environment. The agents can then be trained through on-policy or off-policy methods with the data generated from interactions with the simulated environment rather than the real environment. As a result, users would not be involved directly during the training stage. However, model-based RL  suffers from the following issues:
\begin{itemize}
    \item The reward estimation of the simulator can be affected by various biases in the training data \cite{huang2020keepingbiasout,chen2020biassurvey}.
    \item The transition between (user) states is dynamic and difficult to model \cite{koren2009timesvd}.
    \item The generalization ability of the constructed simulator is not well justified \cite{arora2017generalization}.
\end{itemize}
\subsection{Prompt and Knowledge Base}
A new learning paradigm namely prompt learning has become an emergent topic in the field of NLP. Different from the widely adopted ``pre-training and fine-tuning''---which first pre-trains the model with tasks like language modeling and then fine-tunes learned parameters for downstream tasks---prompt learning aims to use the pre-trained model (which was pre-trained on a large training corpus) as a knowledge base and then formulates the downstream tasks as a prompt \cite{liu2021promptsurvey}. 
Since in the historical offline data we can compute the exact cumulative reward at each interaction step\footnote{In this work, we do not consider the case of delayed rewards.}, the historical offline data can be formulated in the following way \{state, cumulative reward\}-->\{observed action\}, which can be intuitively interpreted as the signal for: \emph{which action should be taken to obtain the cumulative reward given the user state}. During inference, we can expect the model to suggest actions which, if taken, should achieve the prompted reward in expectation, as shown in Figure \ref{fig:prl}. Such a learning paradigm enables us to train the reward-driven RL recommendation agent in a much simpler supervised fashion. Note that, in this work, we exploit the concept of prompt to inspire our approach, but we don't investigate the complex prompt generation methods of NLP. The prompt used in this paper is the pair of state and cumulative reward. We leave more advanced prompt generation for recommendation as future work.

\section{Methodology}
In this section, we describe the detailed training and inference procedures for next item recommendation with PRL.
\subsection{PRL Training}
Training PRL consists of prompt generation, prompt encoding, and supervised attentive learning as shown in Figure \ref{fig:prltrain}.

\subsubsection{Prompt Generation} Prompt generation aims to formulate the offline training data as a knowledge template which tells us which observed action $x_{t+1}$ should be taken if we want to get the cumulative reward $R_t$ with $x_{1:t}$ as the previous user-item interactions. In the offline training data, the cumulative reward $R_t$ at each interaction step $t$ of a session can be exactly computed as:
\begin{equation}
    \label{eq:cumulative_reward}
    R_t=\sum_{t'=t}^{|\tau|}\lambda^{t'}r_{t'},
\end{equation}
where $|\tau|$ denotes the total steps of the interaction session.
For each interaction session in the offline training data, we can inefficiently compute the cumulative reward $R_t$ at every step as shown in Eq.(\ref{eq:cumulative_reward}) with a total time complexity of $O(|\tau|^2)$.
A more efficient solution is to compute $R_1$ firstly and then $R_{t+1}$ can be computed from $R_{t}$ recursively by decreasing the reward $\lambda^{t}r_t$.
With this recursive procedure, we can compute these rewards in $O(|\tau|)$ instead.
Algorithm 1 shows the detailed procedure to reformulate the offline training sequences $\mathcal{D}_s$ as a prompt-based training set $\mathcal{D}_p$.
\begin{algorithm}[b]
 \label{alg:promptgeneration}
 \caption{Prompt generation from offline training data}
 	\begin{algorithmic}[1]
        \renewcommand{\algorithmicrequire}{\textbf{Input:}} 
        \renewcommand{\algorithmicensure}{\textbf{Output:}}
        \Require
        user-item interaction sequence set $\mathcal{D}_s$, reward settings
        \Ensure
        prompt-based training set $\mathcal{D}_p$
        \Repeat 
            \State Sample an interaction sequence $x_{1:T}$ from $\mathcal{D}_s$
            \State Compute $R_1$ according to Eq.(\ref{eq:cumulative_reward})
            \For{t=$1:T-1$}
            \State $\mathcal{D}_p$.append(\{$R_t$,$x_{1:t}$,$t$\},$x_{t+1}$)
            \State $R_{t+1}=R_t-\lambda^tr_t$ 
            \EndFor
            \State $\mathcal{D}_s$.remove($x_{1:T}$)
        \Until $\mathcal{D}_s=\emptyset$
        \State return $\mathcal{D}_p$
 	\end{algorithmic}
 \end{algorithm}
\subsubsection{Prompt Encoding} Prompt encoding aims to use deep neural networks to map the generated prompt into latent representations. For the interaction sequence $x_{1:t}$, plenty of research has been proposed to capture the sequential signals, such as recurrent neural network (RNN)-based methods \cite{gru4rec}, convolutional neural network (CNN)-based methods \cite{caser-rec,nextitnet}, and attention-based methods \cite{SASRec,sun2019bert4rec}. We will give a more detailed description in section \ref{related-work}.
The proposed PRL acts as a learning paradigm and all of these methods can be used as the sequential model shown in Figure \ref{fig:prltrain}. 

Take the gated recurrent units (GRU) \cite{gru4rec} as an example, we first embed each item $x_i$ into a dense representation $\mathbf{x}_i\in\mathbb{R}^d$, where $d$ denotes the embedding size. This can be done through a simple embedding table lookup operation. Then, the hidden state $\mathbf{s}_t$ for a given sequence of $x_{1:t}$ is defined as:
\begin{equation}
    \label{eq:gru}
    \begin{aligned}
\mathbf{s}_t&=(1-\mathbf{z}_t)\mathbf{s}_{t-1}+\mathbf{z}_t\mathbf{\hat{s}_t} \\
\mathbf{z}_t&=\sigma(\mathbf{W}_z\mathbf{x}_t+\mathbf{U}_z\mathbf{s}_{t-1})\\
\mathbf{\hat{s}_t}=&tanh(\mathbf{W}_s\mathbf{x}_t+\mathbf{U}_s(\mathbf{g}_t\odot\mathbf{s}_{t-1}))\\
\mathbf{g}_t&=\sigma(\mathbf{W}_g\mathbf{x}_t+\mathbf{U}_g\mathbf{s}_{t-1}),
\end{aligned}
\end{equation}
where $\sigma$ denotes the sigmoid function and $\odot$ is element-wise product. $\mathbf{W}_z, \mathbf{U}_z,\mathbf{W}_s,\mathbf{U}_s, \mathbf{W}_g, \mathbf{U}_g \in \mathbb{R}^{d\times d}$ are trainable parameters. In our experiments, we use four renowned sequential models to encode $x_{1:t}$. This allows us to verify the effectiveness and generalization ability of the proposed PRL. We don't elaborate on the details of all models, though, since this is not the key point of this work.

The representation for cumulative reward $R_t$ is defined as:
\begin{equation}
    \mathbf{e}_{R_t}=R_t\cdot \mathbf{e}_r,
\end{equation}
where $\mathbf{e}_r \in \mathbb{R}^d$ is a trainable reward embedding. Another solution would be using different reward
embeddings for discretized rewards. Besides, we also maintain a trainable embedding table $\mathbf{H}_T \in \mathbb{R}^{T\times d}$ to encode the step information. The final representation for the prompt \{$R_t$,$x_{1:t}$,$t$\} is formulated as:
\begin{equation}
\label{eq:pt}
    \mathbf{P}_t=[\mathbf{e}_{R_t},\mathbf{s}_t,\mathbf{h}_t]\in \mathbb{R}^{3 \times d},
\end{equation}
where [$\cdot$] denotes the stack operation.
\subsubsection{Supervised Attentive Learning} Given the encoded prompt representation $\mathbf{P}_t$, we need a model to learn to map the signal between $\mathbf{P}_t$ and observed action $a_t$ (i.e., $x_{t+1}$). Self-attention \cite{Transformer} has been widely adopted in the field of NLP and has demonstrated impressive model capability. Recently, there are also works \cite{parisotto2019stabilizing,janner2021trajectory, chen2021decision} attempting to introduce self-attention to RL.
Inspired by this research, we propose to use a self-attentive block to learn the mapping signal.
The dot-product based attention \cite{Transformer} is formulated as:
\begin{equation}
\label{eq:attention}
    \mathrm{Attention}(\mathbf{Q},\mathbf{K},\mathbf{V})=\mathrm{softmax}\left(\frac{\mathbf{Q}\mathbf{K}^T}{\sqrt{d}}\right)\mathbf{V},
\end{equation}
where $\mathbf{Q,K,V}$ denote the queries, keys and values, respectively. 
The attention computes a weighted addition of values according to the importance weights computed through the correlations between the query and the key \cite{SASRec,Transformer}. The scale factor $\sqrt{d}$ is used to normalize the computed values to avoid large inner products, especially when the dimension of the representations is large \cite{SASRec,Transformer}.

Self-attention uses the same objects as queries, keys, and values. In the proposed PRL, we convert $\mathbf{P}_t$ to three representations, using linear projections, and then feed them to the attention layer. The residual connection \cite{he2016deep} is introduced to incorporate the original $\mathbf{P}_t$ information. The final prompt representation $\mathbf{\tilde{P}}_t$ is defined as:
\begin{equation}
\label{eq:selfattentionprl}
    \mathbf{\tilde{P}}_t=\mathbf{P}_t + \mathrm{Attention}(\mathbf{W}_q\mathbf{P}_t,\mathbf{W}_k\mathbf{P}_t,\mathbf{W}_v\mathbf{P}_t),
\end{equation}
where $\mathbf{W}_q, \mathbf{W}_k, \mathbf{W}_v \in \mathbb{R}^{d\times d}$ are trainable parameters.
To avoid overfitting and enable a more stable learning without vanishing or exploding gradient issues, we include optional dropout layers and layer normalization \cite{ba2016layernorm}.

Three attentive representations can then be extracted from $\mathbf{\tilde{P}}_t$:\looseness=-1
\begin{equation}
\label{eq:untack}
    \mathbf{\tilde{e}}_{R_t},\mathbf{\tilde{s}}_t,\mathbf{\tilde{h}}_t=unstack(\mathbf{\tilde{P}}_t).
\end{equation}
We feed the attentive state representation $\mathbf{\tilde{s}}_t$ to a fully connected layer to compute the classification logits on the candidate actions.
\begin{equation}
\label{eq:fc}
    [y_1,y_2,...y_n]=\delta(\mathbf{W}_i\mathbf{\tilde{s}}_t+\mathbf{b}),
\end{equation}
where $\delta$ denotes the activation function and $n$ is the number of candidate actions. $\mathbf{W}_i \in \mathbb{R}^{n\times d}$ can be seen as another trainable item embedding matrix and $\mathbf{b} \in \mathbb{R}^n$ is a bias vector.

Actor-Critic \cite{konda2000actor-critic,xin2020self} methods have achieved excellent results in recent research. The key idea is to use the predicted Q-values from the critic to re-weight the actor so that actions with higher cumulative reward would have more effect in the training stage. In PRL we re-weight the training samples with the immediate reward $r_t$ for a more stable training. The weighted supervised cross-entropy loss is defined as:
\begin{equation}
	\label{eq:supervised-loss}
	L=-r_t\sum_{i=1}^n Y_i \log(p_i), \text{where } p_i=\frac{e^{y_i}}{\sum_{i'=1}^ne^{y_{i'}}}.
\end{equation}
$Y_i$ is an indicator function which is defined as $Y_i=1$ if the user interacted with the $i$-th item in the next timestamp. Otherwise, $Y_i=0$. Algorithm 2 shows a detailed training procedure of PRL.

\begin{algorithm}[b]
 \label{alg:prltraining}
 \caption{Overall Training procedure of PRL}
 	\begin{algorithmic}[1]
        \renewcommand{\algorithmicrequire}{\textbf{Input:}} 
        \renewcommand{\algorithmicensure}{\textbf{Output:}}
        \Require
        user-item interaction sequence set $\mathcal{D}_s$, reward settings
        \Ensure
        all parameters in the learning space $\theta$
        \State Initialize all trainable parameters
        \State Generate $\mathcal{D}_p$ according to Algorithm 1
        \Repeat 
            \State Draw a mini-batch of \{$R_t$,$x_{1:t}$,$t$\},$x_{t+1}$ from $\mathcal{D}_p$
            \State Compute $\mathbf{P}_t$ according to Eq.(\ref{eq:gru})-Eq.(\ref{eq:pt})
            \State Compute $\mathbf{\tilde{s}}_t$ according to Eq.(\ref{eq:selfattentionprl})-Eq.(\ref{eq:untack})
            \State Compute loss function $L$ according to Eq.(\ref{eq:fc})-Eq.(\ref{eq:supervised-loss})
            \For {each parameter $\vartheta \in \theta$}
                \State {Compute $\partial L$/$\partial \vartheta$ on the mini-batch by back-propagation}
                \State {Update $\vartheta \leftarrow \vartheta- \eta \cdot \partial L/\partial \vartheta$}
            \EndFor
        \Until converge
        \State return all parameters in $\theta$
 	\end{algorithmic}
 \end{algorithm}
\subsection{PRL Inference}
In training PRL, the cumulative reward can be computed from the offline data. At inference time, however, we need to provide the model with how much reward we want to obtain; the agent can then adjust its actions conditioning on the prompted reward. For an interaction step $t$, a concise prompt reward can be set as the average cumulative reward of the offline training data at this step. To make the model more flexible, we extend the prompt reward for PRL inference as:
\begin{equation}
    \label{eq:rewardinference}
    \tilde{R}_t=\mathcal{N}(\mu,\epsilon^2)\times \bar{R}_t,
\end{equation}
where $\bar{R}_t$ denotes the average cumulative reward of step $t$ in the training data. $\mathcal{N}(\mu,\epsilon^2)$ is a Gaussian distribution with $\mu$ as the mean and $\epsilon$ as the standard deviation. 
There are also various inference reward settings (e.g., according to the maximum cumulative reward in the training data).
For offline inference and evaluation, we can prompt the model with such expected reward at each timestamp. Besides, for online inference, the proposed PRL can also support sequence-wise recommendation generation through promoting the model with the expected cumulative reward at the beginning (i.e., the first timestamp) and then decreasing the obtained reward according to the real user feedback.
A more desired setting could be that the prompt inference reward can be automatically adjusted given the user state. 
We leave more advanced inference reward settings for future work. 
\section{Experiments}
In this section, we perform experiments\footnote{Codes and data can be accessed in \url{https://drive.google.com/file/d/1Mm5SxNDkdfdUpdhnosQAd9YBoi3QDIp_/view?usp=sharing}} on two e-commerce  datasets to verify the effectiveness of the PRL learning paradigm. We aim to answer the following research questions:

\textbf{RQ1:} How does PRL perform when instantiated with different sequential recommendation models?
	
\textbf{RQ2:} What is the effect of the supervised attentive learning, including the self-attentive block and the weighted loss function?

\textbf{RQ3:} How do the prompt reward settings in the inference stage affect the PRL performance?

\subsection{Experimental Settings}
\subsubsection{Datasets}
Experiments are conducted on two public accessible datasets: Challenge15\footnote{\url{https://recsys.acm.org/recsys15/challenge/}} and RetailRocket\footnote{\url{https://www.kaggle.com/retailrocket/ecommerce-dataset}}. 

\textbf{Challenge15.} This dataset comes from the RecSys Challange 2015. In it, each user--item interaction session contains a sequence of user click or purchases behaviours.  Sessions whose length are shorter than 3 items or longer than 50 are removed. Then 200k sessions are randomly sampled to obtain a dataset containing 1,110,965 clicks and 43,946 purchases upon 26,702 items. 

\textbf{RetailRocket.} This dataset contains sequential data of user's behaviour in a e-commerce website; where users view and add items to a shopping cart. For simplicity, we treat views as clicks and adding to a cart as a purchase. Items which are interacted less than 3 times are removed. Sequences whose length is shorter than 3 or longer than 50 items are also removed. The processed dataset contains 1,176,680 clicks and 57,269 purchases over 70,852 items.
Table \ref{Datasets} presents these datasets' detailed statistics.
\begin{table}
    \centering
    \begin{threeparttable}
    \caption{Dataset statistics.}
    \label{Datasets}
    \begin{tabular}{p{1.5cm}p{2.0cm}<{\centering}p{2.0cm}<{\centering}}
    \toprule
    Dataset&Challenge15 & RetailRocket\cr
    \midrule
    \#sequences&200,000&195,523\cr
    \#items&26,702&70,852\cr
    \#clicks&1,110,965&1,176,680\cr
    \#purchase&43,946&57,269\cr
    \bottomrule
  \end{tabular}
    \end{threeparttable}
\end{table}

\subsubsection{Evaluation protocols}
PRL implements offline training of a RL-based recommendation agent. As a result, the experiments focus is offline evaluation of the PRL agent. The ratio of training, validation, and test set is 8:1:1. We use the same data splits as \cite{xin2020self}. For validation and testing, the evaluation is performed by providing the agent with previous user-item interaction sequences and the generated prompt reward from Eq.(\ref{eq:rewardinference}). Then we check the rank of the ground-truth action (i.e., interacted items) for the next step. The ranking is performed among the whole item set. Each experiment is repeated 3 times, and the average performance is reported.

\begin{table*}
    \centering
    \begin{threeparttable}
    \caption{Top-$k$ recommendation performance comparison of different models ($k=5, 10, 20$) on the Challenge15 dataset. NG is short for NDCG. Boldface denotes the highest score. $*$ denotes the significance $p$-value < 0.1 compared with the best baseline which is marked with \underline{\hspace{0.25cm}}. The values for normal training, SQN and SAC come from \cite{xin2020self}, since we use the same data splits and hyperparameter settings.}\vspace{-5pt}
    \label{comparison between different models on RC15}
    \begin{tabular}{p{1.7cm}p{0.85cm}<{\centering}p{0.85cm}<{\centering}p{0.85cm}<{\centering}p{0.85cm}<{\centering}p{0.85cm}<{\centering}p{0.85cm}<{\centering}p{0.85cm}<{\centering}p{0.85cm}<{\centering}p{0.85cm}<{\centering}p{0.85cm}<{\centering}p{0.85cm}<{\centering}p{0.85cm}<{\centering}}
    \toprule
    \multirow{2}{*}{Models}&\multicolumn{6}{c}{purchase}&\multicolumn{6}{c}{click}\cr
    \cmidrule(lr){2-7} \cmidrule(lr){8-13}
    &HR@5&NG@5&HR@10&NG@10&HR@20&NG@20&HR@5&NG@5&HR@10&NG@10&HR@20&NG@20\cr
    \midrule
    GRU    &0.3994&0.2824 &0.5183&0.3204 &0.6067&0.3429 &0.2876&0.1982 &0.3793&0.2279 &0.4581&0.2478\cr
    GRU-SQN&$0.4228$&$0.3016$&$0.5333$&$0.3376$&$0.6233$&$0.3605$&$\underline{0.3020}$ &$\underline{\textbf{0.2093}}$ &$\underline{0.3946}$&$\underline{0.2394}$&$\underline{0.4741}$&$\underline{0.2587}$\cr
    GRU-SAC&$\underline{0.4394}$&$\underline{0.3154}$&$\underline{0.5525}$&$\underline{0.3521}$&$\underline{0.6378}$&$\underline{0.3739}$&0.2863 &0.1985 &0.3764 &0.2277 &0.4541&0.2474\cr
    GRU-PRL&$\textbf{0.4514}^*$&\textbf{0.3214}&$\textbf{0.5673}^*$&\textbf{0.3593}&$\textbf{0.6525}^*$&$\textbf{0.3809}^*$&\textbf{0.3027}&0.2086&\textbf{0.3967}&\textbf{0.2398}&\textbf{0.4755}&\textbf{0.2598}\\ \hline
    Caser& 0.4475&0.3211 &0.5559&0.3565&0.6393&0.3775&0.2728 &0.1896&0.3593&0.2177&0.4371&0.2372\cr
    Caser-SQN&$0.4553$&$0.3302$&$0.5637$&$0.3653$&$0.6417$&$0.3862$&\underline{0.2742}&\underline{0.1909} &\underline{0.3613}&\underline{0.2192}&\underline{0.4381}&\underline{0.2386}\cr
    Caser-SAC&$\underline{0.4866}$&$\underline{0.3527}$&$\underline{0.5914}$&$\underline{0.3868}$&$\underline{0.6689}$&$\underline{0.4065}$&0.2726&0.1894&0.3580&0.2171&0.4340&0.2362\cr
    Caser-PRL&$\textbf{0.4938}^*$&\textbf{0.3555}&$\textbf{0.6052}^*$&$\textbf{0.3920}^*$&$\textbf{0.6914}^*$&$\textbf{0.4138}^*$&$\textbf{0.3074}^*$&$\textbf{0.2121}^*$&$\textbf{0.4028}^*$&$\textbf{0.2431}^*$&$\textbf{0.4838}^*$&$\textbf{0.2637}^*$\\ \hline
    NItNet&0.3632&0.2547&0.4716&0.2900&0.5558&0.3114&0.2950&0.2030&0.3885&0.2332&0.4684&0.2535\cr
    NItNet-SQN&$0.3845$&$0.2736$&$0.4945$&$0.3094$&$\underline{0.5766}$&$0.3302$&$\underline{0.3091}$ &$\underline{0.2137}$ &$\underline{0.4037}$ &$\underline{0.2442}$&$\underline{0.4835}$&$\underline{0.2645}$\cr
    NItNet-SAC&$\underline{0.3914}$&$\underline{0.2813}$&$\underline{0.4964}$&$\underline{0.3155}$&$0.5763$&$\underline{0.3357}$&$0.2977$&$0.2055$&0.3906&$0.2357$&0.4693&$0.2557$\cr
    NItNet-PRL&$\textbf{0.4295}^*$&$\textbf{0.3098}^*$&$\textbf{0.5405}^*$&$\textbf{0.3460}^*$&$\textbf{0.6351}^*$&$\textbf{0.3701}^*$&$\textbf{0.3280}^*$&$\textbf{0.2273}^*$&$\textbf{0.4248}^*$&$\textbf{0.2588}^*$&$\textbf{0.5028}^*$&$\textbf{0.2786}^*$\\ \hline
    SASRec& 0.4228&0.2938 &0.5418&0.3326&0.6329&0.3558&0.3187&0.2200&0.4164&0.2515&0.4974&0.2720\cr
    SASRec-SQN&0.4336&$0.3067$&0.5505&$0.3435$& $0.6442$&$0.3674$ &$\textbf{\underline{0.3272}}$&$\textbf{\underline{0.2263}}$&$\textbf{\underline{0.4255}}$&$\textbf{\underline{0.2580}}$&$\textbf{\underline{0.5066}}$&$\textbf{\underline{0.2786}}$\cr
    SASRec-SAC&$\underline{0.4540}$&$\underline{0.3246}$&$\underline{0.5701}$&$\underline{0.3623}$&$\underline{0.6576}$& $\underline{0.3846}$&0.3130&0.2161&0.4114&0.2480&0.4945&0.2691\cr
    SASRec-PRL&$\textbf{0.4681}^*$&$\textbf{0.3360}^*$&$\textbf{0.5927}^*$&$\textbf{0.3768}^*$&$\textbf{0.6893}^*$&$\textbf{0.4013}^*$&0.3239&$0.2246$&$0.4219$&$0.2565$&$0.5029$&$0.2770$\\
    \bottomrule
    \end{tabular}
    \end{threeparttable}
    \vspace{0.2cm}
\end{table*}

\begin{table*}
    \centering
    \begin{threeparttable}
    \caption{Top-$k$ recommendation performance comparison of different models ($k=5, 10, 20$) on the RetailRocket dataset. NG is short for NDCG. Boldface denotes the highest score. $*$ denotes the significance $p$-value < 0.1 compared with the best baseline which is marked with \underline{\hspace{0.25cm}}. The values for normal training, SQN and SAC come from \cite{xin2020self}, since we use the same data splits and hyperparameter settings.}\vspace{-5pt}
    \label{comparison between different models on RetailRocket}
    \begin{tabular}{p{1.7cm}p{0.85cm}<{\centering}p{0.85cm}<{\centering}p{0.85cm}<{\centering}p{0.85cm}<{\centering}p{0.85cm}<{\centering}p{0.85cm}<{\centering}p{0.85cm}<{\centering}p{0.85cm}<{\centering}p{0.85cm}<{\centering}p{0.85cm}<{\centering}p{0.85cm}<{\centering}p{0.85cm}<{\centering}}
    \toprule
    \multirow{2}{*}{Models}&\multicolumn{6}{c}{purchase}&\multicolumn{6}{c}{click}\cr
    \cmidrule(lr){2-7} \cmidrule(lr){8-13}
    &HR@5&NG@5&HR@10&NG@10&HR@20&NG@20&HR@5&NG@5&HR@10&NG@10&HR@20&NG@20\cr
    \midrule
    GRU    &0.4608 &0.3834&0.5107 &0.3995&0.5564&0.4111&0.2233&0.1735&0.2673&0.1878&0.3082&0.1981\cr
    GRU-SQN&$\underline{0.5069}$&$0.4130$&$\underline{0.5589}$&$0.4289$&$\underline{0.5946}$&$0.4392$&$\underline{0.2487}$ &$\underline{0.1939}$ &$\underline{0.2967}$&$\underline{0.2094}$&$\underline{0.3406}$&$\underline{0.2205}$\cr
    GRU-SAC&$0.4942$&$\underline{0.4179}$&$0.5464$&$\underline{0.4341}$&$0.5870$&$\underline{0.4428}$&$0.2451$&$0.1924$&$0.2930$ &$0.2074$&$0.3371$&$0.2186$\cr
    GRU-PRL&$\textbf{0.5486}^*$&$\textbf{0.4640}^*$&$\textbf{0.5972}^*$&$\textbf{0.4798}^*$&$\textbf{0.6284}^*$&$\textbf{0.4879}^*$&$\textbf{0.2805}^*$&$\textbf{0.2165}^*$&$\textbf{0.3325}^*$&$\textbf{0.2336}^*$&$\textbf{0.3821}^*$&$\textbf{0.2462}^*$\\ \hline
    Caser& 0.3491&0.2935 &0.3857&0.3053&0.4198&0.3141&0.1966&0.1566&0.2302&0.1675&0.2628&0.1758\cr
    Caser-SQN&$0.3674$&$0.3089$&$0.4050$&$0.3210$&$0.4409$&$0.3301$&$0.2089$&$0.1661$ &$0.2454$&$0.1778$&$0.2803$&$0.1867$\cr
    Caser-SAC&$\underline{0.3871}$&$\underline{0.3234}$&$\underline{0.4336}$&$\underline{0.3386}$&$\underline{0.4763}$&$\underline{0.3494}$&$\underline{0.2206}$&$\underline{0.1732}$&$\underline{0.2617}$&$\underline{0.1865}$&$\underline{0.2999}$&$\underline{0.1961}$\cr
    Caser-PRL&$\textbf{0.5277}^*$&$\textbf{0.4403}^*$&$\textbf{0.5742}^*$&$\textbf{0.4554}^*$&$\textbf{0.6124}^*$&$\textbf{0.4653}^*$&$\textbf{0.2770}^*$&$\textbf{0.2158}^*$&$\textbf{0.3296}^*$&$\textbf{0.2328}^*$&$\textbf{0.3774}^*$&$\textbf{0.2450}^*$\\ \hline
    NItNet&0.5630&0.4630&0.6127&0.4792&0.6477&0.4881&0.2495&0.1906&0.2990&0.2067&0.3419&0.2175\cr
    NItNet-SQN&$0.5895$&$0.4860$&$\textbf{\underline{0.6403}}$&$0.5026$&$\textbf{\underline{0.6766}}$&$0.5118$&$\underline{0.2610}$ &$\underline{0.1982}$ &$\underline{0.3129}$ &$\underline{0.2150}$&$\underline{0.3586}$&$\underline{0.2266}$\cr
    NItNet-SAC&$\underline{0.5895}$&$\underline{0.4985}$&$0.6358$&$\underline{0.5162}$&$0.6657$&$\underline{0.5243}$&$0.2529$&$0.1964$&$0.3010$&$0.2119$&$0.3458$&$0.2233$\cr
    NItNet-PRL&$\textbf{0.5976}^*$&$\textbf{0.5095}^*$&$0.6386$&$\textbf{0.5229}$&$0.6674$&$\textbf{0.5302}$&$\textbf{0.2812}^*$&$\textbf{0.2180}^*$&$\textbf{0.3343}^*$&$\textbf{0.2353}^*$&$\textbf{0.3825}^*$&$\textbf{0.2475}^*$\\ \hline
    SASRec&0.5267&0.4298&0.5916&0.4510&0.6341&0.4618&0.2541&0.1931&0.3085&0.2107&0.3570&0.2230\cr
    SASRec-SQN&$\textbf{\underline{0.5681}}$&$0.4617$&$\textbf{\underline{0.6203}}$&$0.4806$&$\textbf{\underline{0.6619}}$&$0.4914$&$\underline{0.2761}$&$\underline{0.2104}$&$\underline{0.3302}$&$\underline{0.2279}$&$\underline{0.3803}$&$\underline{0.2406}$\cr
    SASRec-SAC&$0.5623$&$\underline{0.4679}$&$0.6127$&$\underline{0.4844}$&$0.6505$& $\underline{0.4940}$&$0.2670$&$0.2056$&$0.3208$&$0.2230$&$0.3701$&$0.2355$\cr
    SASRec-PRL&0.5612&$\textbf{0.4737}^*$&0.6127&$\textbf{0.4905}^*$&0.6564&$\textbf{0.5016}^*$&$\textbf{0.2867}^*$&$\textbf{0.2201}^*$&$\textbf{0.3415}^*$&$\textbf{0.2379}^*$&$\textbf{0.3952}^*$&$\textbf{0.2515}^*$\\
    \bottomrule
    \end{tabular}
    \end{threeparttable}
    \vspace{0.2cm}
\end{table*}

For the main results, the recommendation performance is measured by hit ratio (HR) and normalized discounted cumulative gain (NDCG). HR@$k$ is a recall-based metric, measuring whether the ground-truth action is in the top-$k$ positions of the recommendation list \cite{xin2020self}. We can define HR for clicks as:
\begin{equation}
	\label{HR_click}
	\mathrm{HR}(\mathtt{click})=\frac{\#\text{hits among clicks}}{\#\text{clicks}}
\end{equation}
HR(purchase) is then defined similarly to HR(click), except that we replace numbers of clicks with purchases \cite{xin2020self}. NDCG is a rank weighted metric which assigns higher scores to top ranked positions in the recommendation list \cite{NDCG}.
\begin{table*}
    \centering
    \begin{threeparttable}
    \caption{Effect of the self-attentive block. Boldface is the highest score. $*$ denotes $p$-value < 0.1 compared with PRL.}
    \label{effect-self-attention}
    \begin{tabular}{p{1.3cm}p{1.5cm}<{\centering}p{0.83cm}<{\centering}p{0.83cm}<{\centering}p{0.83cm}<{\centering}p{0.83cm}<{\centering}p{0.83cm}<{\centering}p{0.83cm}<{\centering}p{0.83cm}<{\centering}p{0.83cm}<{\centering}p{0.83cm}<{\centering}p{0.83cm}<{\centering}p{0.83cm}<{\centering}p{0.83cm}<{\centering}}
    \toprule
    &\multirow{2}{*}{Methods}&\multicolumn{6}{c}{purchase}&\multicolumn{6}{c}{click}\cr
    \cmidrule(lr){3-8} \cmidrule(lr){9-14}
    &&HR@5&NG@5&HR@10&NG@10&HR@20&NG@20&HR@5&NG@5&HR@10&NG@10&HR@20&NG@20\cr
    \midrule
    \multirow{3}{*}{Challenge15}&PRL-mean&$0.4389^*$&$0.3128^*$&$0.5477^*$&$0.3483^*$&$0.6302^*$&$0.3692^*$&$0.2914^*$&$0.2028^*$&$0.3783^*$&$0.2310^*$&$0.4520^*$&$0.2497^*$\cr
    &PRL-MLP&0.4481&0.3198&0.5618&0.3568&0.6504&\textbf{0.3812}&$0.2840^*$&$0.1962^*$&$0.3761^*$&$0.2261^*$&$0.4584^*$&$0.2469^*$\cr
    &PRL&$\textbf{0.4514}$&\textbf{0.3214}&$\textbf{0.5673}$&\textbf{0.3593}&$\textbf{0.6525}$&$0.3809$&\textbf{0.3027}&\textbf{0.2086}&\textbf{0.3967}&\textbf{0.2398}&\textbf{0.4755}&\textbf{0.2598}\cr
    \midrule
    \multirow{3}{*}{RetailRocket}
    &PRL-mean&$0.5176^*$&$0.4480^*$&$0.5591^*$&$0.4615^*$&$0.5924^*$&$0.4699^*$&$0.2469^*$&$0.1940^*$&$0.2907^*$&$0.2082^*$&$0.3337^*$&$0.2191^*$\cr
    &PRL-MLP&$0.4890^*$&$0.4090^*$&$0.5373^*$&$0.4247^*$&$0.5817^*$&$0.4360^*$&$0.2439^*$&$0.1899^*$&$0.2912^*$&$0.2052^*$&$0.3362^*$&$0.2166^*$\cr
    &PRL&$\textbf{0.5486}$&$\textbf{0.4640}$&$\textbf{0.5972}$&$\textbf{0.4798}$&$\textbf{0.6284}$&$\textbf{0.4879}$&$\textbf{0.2805}$&$\textbf{0.2165}$&$\textbf{0.3325}$&$\textbf{0.2336}$&$\textbf{0.3821}$&$\textbf{0.2462}$\cr
    \bottomrule
    \end{tabular}
    \end{threeparttable}
\end{table*}
\subsubsection{Baselines}
We instantiate PRL with four renowned deep learning-based sequential recommendation models, including RNN-based models, CNN-based models, and attention-based models to verify the generalization ability of the proposed PRL.
\begin{itemize}
    \item GRU \cite{gru4rec}: This method utilizes a GRU to model user--item interactions. The hidden state of the final timestamp is regarded as the environment state, as shown in Eq. (\ref{eq:gru}).
	\item Caser \cite{caser-rec}: This is a CNN-based method which applies convolutions on the item embedding sequence. Caser is effective at capturing skipping signals between interactions.
	\item NItNet \cite{nextitnet}: This method uses a dilated CNN to enlarge the receptive field to learn long sequences. Besides, residual connections are introduced to increase the network depth. NItNet achieves good performances with high efficiency.
	\item SASRec \cite{SASRec}: This baseline is attention-based and uses the Transformer \cite{Transformer} decoder. The output of the Transformer is treated as the state for the previous sequence. 
\end{itemize}
Each model is trained with the following approaches:
\begin{itemize}[leftmargin=*]
  \item Normal: Train the model with the normal cross-entropy loss.
  \item SQN~\cite{xin2020self}: Self-supervised Q-learning is a recently proposed offline RL learning method which combines supervised learning with Q-learning through a shared base model.
  \item SAC~\cite{xin2020self}: Self-supervised actor-critic further extends SQN by using the Q-learning part as a critic to re-weight the supervised learning-based actor.
  \item PRL: Our proposed method.
\end{itemize}

\subsubsection{Parameter settings}
On our experiments using both datasets, we limit the model's input to only use the last 10 interacted items at a time, i.e. using only $x_{t-10:t}$ as our model's input.
For sequences whose lengths are less than 10, we pad these sequences with a padding token.
The Adam optimizer \cite{kingma2014adam} is used to train all models, with batches of size 256. The learning rate is set as 0.01 for RC15 and 0.005 for RetailRocket. For a fair comparison, the item embedding size is set as 64 for all models and all training methods. For GRU, the size of the hidden state is set as 64. For Caser, we use 1 vertical convolution filter and 16 horizontal filters whose heights are set from \{2,3,4\} according to the original paper \cite{caser-rec}. For NextItNet, we use the code published by its authors and keep the settings unchanged. For SASRec, the number of heads in self-attention is set to 1, following the original paper \cite{SASRec}. 
The drop-out ratio is tuned among [0,0.1,0.2,0.3,0.4,0.5] since we observed that continuing increasing the drop ratio would affect the model performance. For SQN and SAC, we use the exact same setting with their original paper \cite{xin2020self}. 
For PRL, the prompt's reward setting used at inference time is set as $\mu=2$ and $\epsilon=0$.
The reward for purchases is set as $r_p=1.0$ and the reward for clicks is set as $r_c=0.2$.
Note that the hyperparameters of recommendation models are kept exactly the same across all training approaches
for a fair comparison. Further, by keeping PRL's hyperparameters constant across our experiments, we show that it can be instantiated with different models without exhaustive hyperparameter refinement.

\subsection{Performance Comparison (RQ1)}

Tables \ref{comparison between different models on RC15} and \ref{comparison between different models on RetailRocket} show a comparison of the top-$k$ recommendation performances on Challenge15 and RetailRocket, respectively. The values for normal training, SQN and SAC come from \cite{xin2020self}, since we use the same data splits and hyperparameter settings.

We observe that on the Challenge15 dataset, the proposed PRL method achieves the best performance in almost all cases except for the click prediction when integrating with the SASRec model, in which case SQN achieves the highest scores. However the performance gap between SASRec-SQN and SASRec-PRL for click prediction is very small and both of the two methods over-perform normal training. The reason of similar click prediction performance between SASRec-SQN and SASRec-PRL could be that the self-attention based SASRec itself is a powerful base model and the candidate item set in the Challenge15 dataset (i.e., 26,702) is relatively small, so both PRL and SQN have been pushed to the similar almost optimal performance. However, we can see that for purchase predictions, SASRec-PRL still achieves significant performance improvement. It demonstrates that PRL effectively improves the offline training performance of RL-based recommender systems.


On the RetailRocket dataset, we can see that PRL also achieves the highest scores in almost all situations. PRL achieves the highest NDCG in all cases. This demonstrates that PRL tends to push the items which have a higher purchase reward to the top ranking positions of the recommendation list. The biggest performance improvement on RetailRocket is achieved when PRL is instantiated with the Caser model. This further verifies the effectiveness and the generalization ability of PRL.

To conclude, PRL consistently and significantly improves the offline learning performance for RL-based recommendation tasks and can be applied for various sequential recommendation models. 
\subsection{Ablation Study (RQ2)}
\subsubsection{Effect of the self-attentive block.}
PRL uses a self-attentive block to map a prompt to a corresponding action. In this section, we conduct experiments to verify the effect of this block. We replace the self-attentive block with mean-pooling (i.e., PRL-mean) or a multi-layer perceptron (MLP) (i.e.,PRL-MLP). Table \ref{effect-self-attention} shows the performance comparison when using GRU as the base sequential model. Results of other models lead to the same conclusion. We can see that PRL with the self-attentive block achieves the best performance with significant improvement. This demonstrates the effectiveness of the self-attentive block. Besides, comparing the results with Table \ref{comparison between different models on RC15} and Table \ref{comparison between different models on RetailRocket}, we can see that PRL-mean and PRL-MLP achieve better performance than the naive GRU. It further demonstrates the involvement of reward prompt-based learning is effective to improve the recommendation performance.
\begin{table*}
    \centering
    \begin{threeparttable}
    \caption{Effect of the weighted loss. 
    PRL-w/o denotes training the agent with PRL but without any re-weighting. PRL-cumu means using the cumulative reward to re-weight the loss. $*$ denotes $p$-value < 0.1 compared with PRL.}
    \label{effectweightedloss}
    \begin{tabular}{p{1.3cm}p{1.5cm}<{\centering}p{0.83cm}<{\centering}p{0.83cm}<{\centering}p{0.83cm}<{\centering}p{0.83cm}<{\centering}p{0.83cm}<{\centering}p{0.83cm}<{\centering}p{0.83cm}<{\centering}p{0.83cm}<{\centering}p{0.83cm}<{\centering}p{0.83cm}<{\centering}p{0.83cm}<{\centering}p{0.83cm}<{\centering}}
    \toprule
    &\multirow{2}{*}{Methods}&\multicolumn{6}{c}{purchase}&\multicolumn{6}{c}{click}\cr
    \cmidrule(lr){3-8} \cmidrule(lr){9-14}
    &&HR@5&NG@5&HR@10&NG@10&HR@20&NG@20&HR@5&NG@5&HR@10&NG@10&HR@20&NG@20\cr
    \midrule
    \multirow{3}{*}{Challenge15}
    &PRL-w/o&$0.4143^*$&$0.2905^*$&$0.5227^*$&$0.3259^*$&$0.6141^*$&$0.3491^*$&$\textbf{0.3135}^*$&$\textbf{0.2166}^*$&$\textbf{0.4080}^*$&$\textbf{0.2473}^*$&$\textbf{0.4889}^*$&$\textbf{0.2678}^*$\cr
    &PRL-cumu&$0.4026^*$&$0.2769^*$&$0.5051^*$&$0.3103^*$&$0.5955^*$&$0.3332^*$&$0.2887^*$&$0.1984^*$&$0.3792^*$&$0.2278^*$&$0.4551^*$&$0.2471^*$\cr
    &PRL&$\textbf{0.4514}$&\textbf{0.3214}&$\textbf{0.5673}$&\textbf{0.3593}&$\textbf{0.6525}$&$\textbf{0.3809}$&0.3027&0.2086&0.3967&0.2398&0.4755&0.2598\cr
    \midrule
    \multirow{3}{*}{RetailRocket}
    &PRL-w/o&$0.5193^*$&$0.4267^*$&$0.5717^*$&$0.4438^*$&$0.6154^*$&$0.4548^*$&$0.2780$&$0.2139$&$\textbf{0.3350}$&$0.2324$&$\textbf{0.3841}$&$0.2448$\cr
    &PRL-cumu&$0.4951^*$&$0.4017^*$&$0.5555^*$&$0.4214^*$&$0.5954^*$&$0.4316^*$&$0.2596^*$&$0.1964^*$&$0.3130^*$&$0.2137^*$&$0.3611^*$&$0.2259^*$\cr
    &PRL&$\textbf{0.5486}$&$\textbf{0.4640}$&$\textbf{0.5972}$&$\textbf{0.4798}$&$\textbf{0.6284}$&$\textbf{0.4879}$&$\textbf{0.2805}$&$\textbf{0.2165}$&$0.3325$&$\textbf{0.2336}$&$0.3821$&$\textbf{0.2462}$\cr
    \bottomrule
    \end{tabular}
    \end{threeparttable}
\end{table*}

\subsubsection{Effect of the weighted loss}
PRL uses the immediate reward $r_t$ to re-weight the supervised training loss so that actions with higher reward would account for larger weights. In this subsection, we conduct experiments to illustrate the effect of this re-weighting schema. We compare the results of PRL without any re-weighting (i.e., PRL-w/o) and PRL re-weighted by the cumulative reward (i.e., PRL-cumu). 
Table \ref{effectweightedloss} shows the performance comparison when using GRU as the base sequential model. 
We can see that on the Challenge15 dataset, PRL-w/o achieves the highest scores for click prediction while PRL achieves the best purchase prediction. On RetailRocket, PRL-w/o and PRL have similar performance for click prediction but PRL performs better for purchase prediction.
The results demonstrate that the re-weighting schema of PRL successfully helps the model to recommend more purchased items with higher immediate reward.
Regarding PRL-cumu, we can see that its performance is worse than PRL-w/o and PRL. The reason could be that the cumulative reward has much higher variance. So directly using the cumulative reward to re-weight the loss cannot boost the agent performance. 
Besides, comparing these results to Tables \ref{comparison between different models on RC15} and \ref{comparison between different models on RetailRocket}, we can see that even PRL-w/o achieves better recommendation performance than the naive GRU. This further demonstrate the effectiveness of the proposed prompt-based learning.

\subsection{Prompt Reward Investigation (RQ3)}

\begin{figure}
    \captionsetup[subfloat]
    {}
    \centering
    \subfloat[Challenge15.]{%
    \label{fig:effectmuRC}
    \includegraphics[width=0.23\textwidth]{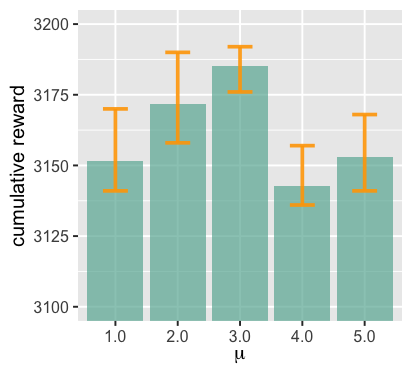}}
    \subfloat[RetailRocket.]{
    \label{fig:effectmuRetail}
    \includegraphics[width=0.23\textwidth]{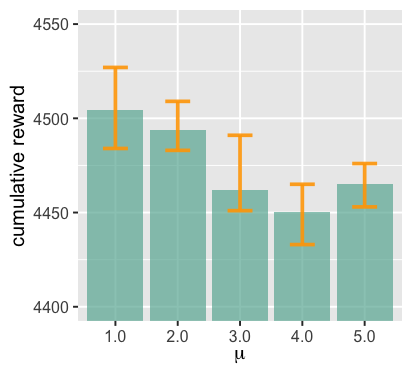}}
    \caption{Effect of the inference reward expectation $\mu$.}
     \label{fig:effectmu}
\end{figure}

\begin{figure}
    \captionsetup[subfloat]
    {}
    \centering
    \subfloat[Challenge15.]{%
    \label{fig:effectstdRC}
    \includegraphics[width=0.24\textwidth]{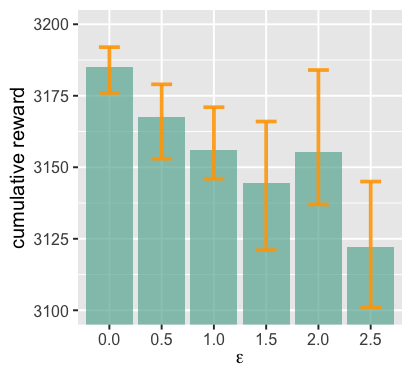}}
    \subfloat[RetailRocket.]{
    \label{fig:effecstdRetail}
    \includegraphics[width=0.24\textwidth]{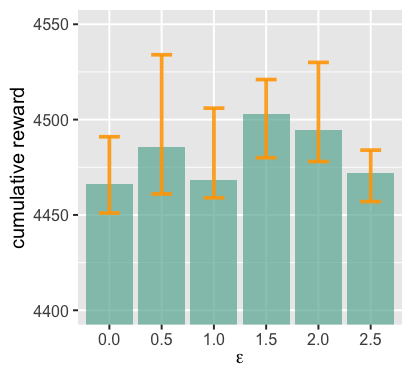}}
    \caption{Effect of the inference reward deviation $\epsilon$.}
    \label{fig:effectstd}
\end{figure}

In this subsection, we conduct experiments to see how the inference reward settings affect the model performance. We report the cumulative reward@1 in the test set, which measures how much cumulative reward we can get from the top-1 position of our recommendation list. Figure \ref{fig:effectmuRC} and Figure \ref{fig:effectmuRetail} show the effect of reward expectation $\mu$ on Challenge15 and RetailRocket, respectively. We can see that on the Challenge15 dataset, the cumulative reward increases at beginning and then decreases. While on the RetailRocket dataset, the cumulative reward keeps decreasing with higher reward expectations. This demonstrates that a larger reward expectation sometimes improves the inference performance, but a reward expectation too large can be harmful. Figures \ref{fig:effectstdRC} and \ref{fig:effecstdRetail} illustrate the effect of the reward deviation $\epsilon$, which can be seen as an exploration factor. We can see that on Challenge15, a large $\epsilon$ reduces the recommendation performance. While on RetailRocket, a larger $\epsilon$ can slightly improve the inference performance. Combined, the results of Figures \ref{fig:effectmu} and \ref{fig:effectstd} suggest that Challenge15's users may prefer items with high reward expectation and little exploration. On the other hand, RetailRocket's users may tend to prefer more exploration. 

\section{Related Work} \label{related-work}
Markov Chain (MC) models \cite{he2016fusingmarkvo,rendle2010factorizingmarkov, koren2009mf} and factorization-based methods \cite{fm,gmf} were widely used for next item recommendation tasks in the past. However, such shallow models cannot effectively capture complex sequential signals \cite{caser-rec,nextitnet}.
Recently, deep learning-based sequential models have been widely investigated for next item recommendation. 
\cite{gru4rec} proposed to model user previous interactions by GRU. 
\cite{caser-rec} and \cite{nextitnet} are based on CNNs.
Besides, \cite{SASRec, sun2019bert4rec} exploited self-attention and Transformer-based architectures \cite{Transformer}.
Generally speaking, all we need is a model $G$ (described in section \ref{Next-item-recommendation}), whose input is a sequence of previous user--item interactions, while the output is a hidden state $\mathbf{s}$ that describes the user's state.
The proposed PRL thus serves as a general learning paradigm, and its model $G$ can be instantiated with any of a diverse pool of sequential models.

RL has been previously applied for recommendation. To perform offline learning from historical data, existing works mainly focus on using IPS score \cite{googlewsdmoffpolicycorrection} and model-based simulation \cite{GAIL, chen2019generativeusermodel, jdkdd19}. 
Besides, \cite{xin2020self} proposed self-supervised reinforcement learning for recommendation; given a standard supervised generative sequential model, they introduce an additional output layer which is trained with standard
Q-learning to bias the model towards the desired reward expectation.
Furthermore, offline RL algorithms are attracting more and more research efforts. \cite{bcq} proposed batch constrained Q-learning, forcing the agent to generate in-distribution actions. \cite{kumar2020conservative} proposed conservative Q-learning to avoid the over-estimation of Q-values. \cite{upsidedown1, upsidedown2} proposed upside-down RL to transform RL into a form of supervised learning. Recently, \cite{janner2021trajectory, chen2021decision} proposed to use Transformers to model the RL problem as a big sequence modeling task. While similar to our proposed prompt-inspired methodology, these methods were not tailored to perform next item recommendation, and could encounter difficulties if applied to a setting with highly dynamic user states \cite{koren2009timesvd} and a large action space \cite{ie2019slateq}.

\section{Conclusion and Future Work}
We propose prompt-based reinforcement learning for the offline training of RL-based next item recommendation agents. We theoretically analyse the offline training challenge when exploiting RL for recommendation. Then we propose to use the historical offline data as a knowledge base and then formulate the recommendation task as a question of which action should be taken if the prompt reward is expected to be achieved under the state observation. The proposed PRL can be trained through a simple supervised fashion. We implement PRL with four renowned sequential recommendation models and conduct experiments on two real-world datasets. Experimental results demonstrate the effectiveness of our proposed method.
Future work includes online tests and more advanced prompt generation. Besides, we are also interested in investigating adaptive prompt reward settings for model inference. 
\begin{acks}
This work is supported by the National Key R\&D Program of China (2020YFB1406704), the Natural Science Foundation of China (61902219, 61972234, 62072279, 62102234), the Key Scientific and Technological Innovation Program of Shandong Province with grant No.2019JZZY010129, the Natural Science Foundation of Shandong Province (ZR2021QF129), the Tencent WeChat Rhino-Bird Focused Research Program (JR-WXG-2021411), the Fundamental Research Funds of Shandong University, the Shandong University multidisciplinary research and innovation team of young scholars (2020QNQT017), and Meituan. All content represents the opinion of the authors, which is not necessarily shared or endorsed by their respective employers and/or sponsors.
\end{acks}

\bibliographystyle{ACM-Reference-Format}
\bibliography{sample-bibliography}

\end{document}